\documentclass{ws-ijbc}
\usepackage{ws-rotating}     
\usepackage{array, multirow}

\input cyracc.def
\font\tencyr=wncyr10
\def\cyr{\tencyr\cyracc}

\begin{document}

\catchline{}{}{}{}{} 

\markboth{J.-M. Ginoux}{The first ``lost'' International Conference on Nonlinear Oscillations (I.C.N.O.)}

\title{The first ``lost'' International Conference on Nonlinear Oscillations (I.C.N.O.)}

\author{JEAN-MARC GINOUX}

\address{Laboratoire {\sc Protee}, I.U.T. de Toulon, Universit\'{e} du Sud, BP 20132, F-83957 La Garde Cedex, France, ginoux@univ-tln.fr, \url{http://ginoux.univ-tln.fr}}

\maketitle

\begin{history}
\received{(to be inserted by publisher)}
\end{history}

\begin{abstract}
From 28 to 30 January 1933 was held at the Institut Henri Poincar\'{e} (Paris) the first International Conference of Nonlinear Oscillations organized at the initiative of the Dutch physicist Balthazar Van der Pol and of the Russian mathematician Nikola\"{i} Dmitrievich Papaleksi. The discovery of this forgotten event, whose virtually no trace remains, was made possible thanks to the report written by Papaleksi at his return in USSR. This document has revealed, one the one hand, the list of participants who included French mathematicians: Alfred Li\'{e}nard, \'{E}lie and Henri Cartan, Henri Abraham, Eug\`{e}ne Bloch, L\'{e}on Brillouin, Yves Rocard, ... and, on the other hand the content of presentations and discussions. The analysis of the minutes of this conference presented here for the first time highlights the role and involvement of the French scientific community in the development of the theory of nonlinear oscillations.
\end{abstract}



\section{Introduction}

In a famous article ``The nonlinear theory of electric oscillations'', published in 1934 in the \textit{Proceedings of the Institute of Radio Engineers} Balthazar Van der Pol \cite[p. 1051]{VdP6} ends his introduction by this sentence:

\begin{quote}

``Although the first researches in connection with our subject date back to
1920 and although the development of this theory has gradually continued
ever since, recent years have shown a considerable increase of activity in
this field by many research workers scattered all over the world, and a
special international conference dedicated solely to the problems arising in
the nonlinear oscillation theory was recently held in Paris, on January
28-30, 1933.''

\end{quote}

Celebrating the centenary of the birth of Papaleksi in 1981, the Russian Vladimir Vasil'evich Migulin \cite[p. 616]{Mig} said that during the ``first International Conference on Nonlinear Oscillations'' which took place in January 1933 in Paris, Nikola\"{i} Dimitrievich Papaleksi presented two papers on research being conducted in the USSR in this domain. Twenty-five years later, the Russian academician Evgenu L'vovich Feinberg \cite[p. 67]{Fein} wrote in a tribute to Papaleksi:

\begin{quote}

``It is not surprising that, when the first international conference on
nonlinear oscillations was convened in Paris in 1932 (among its participants
were such pioneers in this field as B. Van der Pol, L. Brillouin, and
others), it was Papaleksi who represented the Moscow school of Mandel'shtam
and Papaleksi, their closest disciples and colleagues Andronov, A. A. Vitt,
Khaikin, and others, reporting on its achievements.''

\end{quote}

With the exception of these three references, this conference does not seem to appear anywhere in the literature. It contains no information concerning the location where it is supposed to have held in Paris, the list of participants, the program or the proceedings. Also, it seemed justified to ask whether this event had really occurred or not. Furthermore, this international conference is mentioned neither in French scientific journals such as \textit{Comptes Rendus} of the French Academy of Sciences of Paris (C.R.A.S.) nor the \textit{Revue g\'{e}n\'{e}rale des sciences pures et appliqu\'{e}es}, nor in newspapers like \textit{Le Figaro} although it was the case for the International Electrical Congress held in Paris in July 1932. Then, while Van der Pol \cite{VdP6} and Migulin \cite{Mig} speak of a conference held in 1933, Feinberg \cite{Fein} claims that this first conference on nonlinear oscillations was decided in 1932. Finally, Migulin and Feinberg say that Van der Pol, Brillouin and Papaleksi were present. Extensive research in the biography and bibliography of the two first failed to find any trace of this conference. However, we have been able to find a publication of Papaleksi entirely devoted to this event. In an article published in Russian in 1934 in the journal \textit{Zeitschrift f\"{u}r Technische Physik}, entitled {\cyr Mezhdunarodnaya neline{\u i}nym konferentsnya}, i.e. \textit{International Conference on nonlinear process}, Papaleksi \cite{Pa1} gave the minutes of that meeting with great details. The translation of this article has allowed us to describe the course of this first international conference on nonlinear oscillations. After, having found again the reference of this conference, we were able to identify other sources alluding to it (See \cite[p. 4]{Andro3} and \cite[vol. 1, p. 30]{Man}). In a biography of Aleksandr' Adol'fovich Witt, Bendrikov and Sidorova \cite[p. 159]{Ben} write:

\begin{quote}

``In January 1933, was held in Paris an International Conference on Nonlinear Oscillations to which attended Van der Pol (Netherlands), Li\'{e}nard, Cartan, Esclangon, Abraham, L. Brillouin, Le Corbeiller (France), and other eminent mathematicians and physicists from different countries [(Bendrikov and Sidorova refer to Papaleksi \cite{Pa1})]. The representatives of the Soviet Union invited to the conference were L. I. Mandel'shtam, N. M. Kryloff, N. Papaleksi, A. A. Andronov and A. A. Witt. However, the only one to take part in the conference was Papaleksi.''

\end{quote}

However, none of the French present at this meeting seems to have made reference to it in any of its publications. The limited impact of this international conference, the very first to be organized in France with a majority of French scientists, in order to specifically examine the problems occurring in nonlinear oscillations, is therefore a source of questions.

\section{Historical context}

Although the ``audion'', i.e. a forerunner of the triode was invented in 1907 by Lee de Forest (1873-1961) it was not until the First World War that its diffusion began to spread for military and commercial reasons. During the summer of 1914, the French engineer Paul Pichon (18??-1929), who had deserted the French Army in 1900 and had emigrated to Germany, went to the U.S.A. on an assignment from his employers, the Telefunken Company of Germany, to gather samples of all the latest wireless equipment he could find and to return to Germany with his samples for assessment. In the course of his tour he visited the Western Electric Company, and was given samples of the latest high-vacuum Audions together with full information on their use. On his way back to Germany at the end of his American tour he traveled by Atlantic liner to Southampton and he found himself in London on 3 August 1914, the very day upon which Germany declared war on France. Considered as a French deserter in France and as an alien in Germany he was in a very bad situation. However, he decided to cross to Calais where he was arrested and brought before the French military authorities, i.e. the Colonel Gustave Ferri\'{e} (1868-1932), the commandant of the French Military Telegraphic Service. This latter immediately submitted the Audion samples that Pichon  had in his suitcases to a panel of eminent physicists comprising Henri Abraham (1868-1943) who was sent to Lyon in order to reproduce and improve such a device. Less than one year after, the French valve known as the Type T.M. (Military Telegraphy) was born. After many tests it became evident that the French valve were vastly superior in every way to the soft-vacuum Round valves and the earlier model Audions in use hitherto.\\

\newpage

On May 1, 1915, Ferri\'{e} asked to Abraham to come back to Paris where he found again his laboratory at the \'{E}cole Normale Sup\'{e}rieure and his colleague Eug\`{e}ne Bloch (1878-1944). Together, they invented \cite{AB1,AB2} at the end of the First World War\footnote{A report classified ``top secret'' for fifty years found again by Ginoux \cite{Gi2} has enabled to show that the multivibrator was invented in 1917 and not in 1919 as attested by the date of the publications of Abraham and Bloch \cite{AB1,AB2}. See E.C.M.R. report n° 2900, December 1917.} the ``multivibrator'': a device consisting of two lamps T.M. producing oscillations very rich in harmonics and used as a frequency divider. According to Van der Pol \cite[p. 987-988]{VdP2} the concept of ``relaxation oscillations'' finds its source in the works of Abraham and Bloch \cite{AB2} on the multivibrator (See Ginoux \& Letellier \cite{GinLet}). At that time the triode oscillator equation had not been completely stated since the oscillation characteristic, i.e. the electromotive force (potential difference) of the triode related to the current by a nonlinear function had not been yet explicitly defined and modeled.\\

On November 17, 1919, the French engineer Andr\'{e} Blondel (1863-1938) presented at the \textit{C.R.A.S.} a note in which he stated a third-order nonlinear differential equation of the oscillations of a triode \cite{B1}. One year later, July 17, 1920, Van der Pol \cite{VdP1} ends an article that will be only published in November and December 1920 and in which he established the embryonic form of the second-order nonlinear differential equation of the triode oscillator, now referred to as ``van der Pol's equation\footnote{About this question of priority see Ginoux \& Letellier \cite{GinLet}.}''.\\

Five years later, the French mathematician \'{E}lie Cartan and his son Henri Cartan \cite{Ca} proved the uniqueness of the periodic solution of an equation completely analogous to that of Van der Pol \cite{VdP1}.\\

In 1926, Van der Pol wrote an article entitled ``On relaxation oscillations'' whom there exists at least four different versions: two in Dutch, one in German and the last and the most famous in English \cite{VdP2}. In this paper Van der Pol \cite{VdP2} proposed a prototypical second-order differential equation in a nondimensionalized form which will be named later ``Van der Pol's equation'' in order to describe a new phenomenon he called \textit{relaxation oscillations}. By using the classical graphical integration method of the nullclines he plotted the periodic solution of its equation. Let's notice that Van der Pol didn't realized that it was a limit cycle according to Poincar\'{e} \cite[p. 261]{P2} and he didn't use this expression till 1930 (See \cite[p. 294]{VdP4}), i.e. till the presentation of the famous note\footnote{It has been stated by Ginoux \cite{Gi2} that the first paper of Andronov on the theory of oscillations and Poincar\'{e}'s limit cycle was originally published in 1928 during the congress of Russian Physicists held in Moscow on August 5-16, 1928. See Andronov \cite{Andro1}. Moreover, according to Boyco \cite[p. 30]{Boy} Andronov's thesis has not been preserved.} by Andronov \cite{Andro2}. Moreover, he didn't seem to be aware of the result of Cartan \cite{Ca}.\\

Starting from a more general equation than that stated by Van der Pol \cite{VdP2} and following the works of Cartan \cite{Ca}, Alfred Li\'{e}nard \cite{Lien1} demonstrated the existence and uniqueness of the periodic solution of his equation. Nevertheless, he didn't made either any connection between the periodic solution he studied and the concept of limit cycle introduced by Poincar\'{e} \cite[p. 261]{P2}.\\

According to the historiography and till recently, Andronov \cite{Andro2} was considered as the first to realize a ``correspondence'' between the periodic solution of an oscillator of the radiophysics such as that of Van der Pol and the concept of limit cycle of Poincar\'{e}. It has been established by Ginoux \textit{et al.} \cite{Gi2} that in a series of ``forgotten'' conferences given in 1908, Poincar\'{e} applied his own concept of limit cycle to state the stability of oscillations maintained in a radiophysics device: the singing arc. Unfortunately this work which has been rediscovered two years ago doesn't seem to have attracted the attention of contemporary researchers of Poincar\'{e} such as Blondel, Cartan, Li\'{e}nard or Van der Pol.

\vspace{0.1in}

From the late 1920s, Van der Pol \cite{VdP3,VdP4} was regularly invited in France to present his works on relaxation oscillations. During his lectures at the \'{E}cole Sup\'{e}rieure d'\'{E}lectricit\'{e} in Paris on 10 and 11 March 1930, Van der Pol \cite[p. 17]{VdP4} recognized then the ``correspondence'' established by Andronov \cite{Andro2} a few months ago.

\newpage

In the beginning of the 1930s, the French engineer Philippe Le Corbeiller (1891-1930), who was assisting Van der Pol during his stay in France, participated with Li\'{e}nard and Van der Pol \cite{VdP5} to the \textit{Third International Congress for Applied Mechanics} which was held in Stockholm from 24 to 29 August 1930. Le Corbeiller \cite[p. 211]{LeCorb1} was then one of the first French scientists to recognize the ``correspondence'' established by Andronov \cite{Andro1,Andro2} a few months ago.\\

Thereafter, he gave many lectures in order to popularize the concept of relaxation oscillations and what Kryloff \& Bogoliouboff \cite{KB1, KB2} have named the \textit{Nonlinear Mechanics}. One of the most famous is that he made at the \textit{Conservatoire National des Arts et M\'{e}tiers} (C.N.A.M.) on 6 and 7 May 1931 \cite{LeCorb2}. In September 1931, Le Corbeiller \cite{LeCorb3} gave a talk in Lausanne in which he asked for the creation of a Theory of Oscillations:

\begin{quote}

``The idea tends to spread that the methods that chance gave rise among telephone engineers must form a body of doctrine, the \textit{Theory of Oscillations}, whose theorems are in principle independent of applications, mechanical, thermodynamic, electrical, chemical, physiological, biological or economic that they might have, but already illustrated by examples from these areas.''

\end{quote}

International scientific community became gradually aware of the necessity and importance of studying nonlinear process and what Mandel'shtam \cite[vol. 3, p. 52]{Man} called the ``nonlinear thinking'' began to emerge. Thus, it was during the \textit{International Electrical Congress}, which took place in Paris from 5 to 12 July 1932 and to which attended Abraham \& Bloch, Brillouin, Blondel, Kryloff \& Bogoliouboff and of course Van der Pol, that the organization of the first \textit{International Conference on Nonlinear Oscillations} was decided.

\section{The location: l'Institut Henri Poincar\'{e}}

As it was reported by Papaleksi \cite[p. 210]{Pa1} ``The conference was held at the Research Institute Henri Poincar\'{e} on January 28, 29 and 30, 1933.'' Unfortunately, research at the Institut Henri Poincar\'{e} (I.H.P.) failed to find any trace of this conference, although it has been established that Nicolas Kryloff and Nicolas Bogoliuboff were invited later in 1935. According to Papaleksi, the January 30, 1933 was devoted to visiting the laboratory of Henri Abraham (1868-1943) at the \'{E}cole Normale Sup\'{e}rieure, where the conference ended. Investigations were also undertaken within this school but without any result.

\begin{figure}[htbp]
\centerline{\includegraphics[width=8.37cm,height=4.84cm]{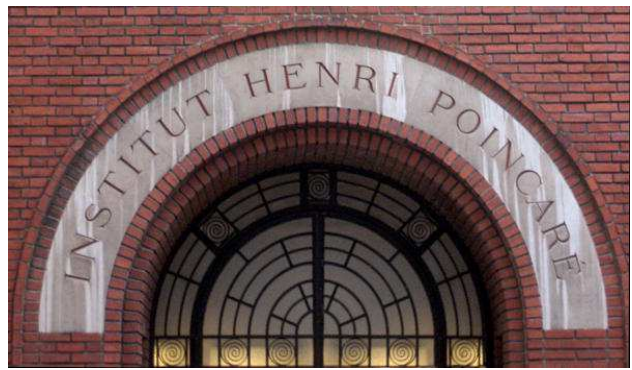}}
\caption[Institut Henri Poincar\'{e}]{Institut Henri Poincar\'{e}, Mosseri \cite[p. 121]{Mos}}
\label{figIHP}
\end{figure}

\section{Listing of participants}

Concerning the participants, the conference was supposed to bring together mathematicians and physicists of international reputation as recalled Papaleksi \cite[p. 210]{Pa1} :

\begin{quote}

``Many people from different countries who are already working in this field and whose cooperation is expected are invited to participate in this conference. Among them we can quote Professor Volterra, which uses mathematical analysis to answer questions from the fluctuations of animal species in the struggle for existence, well-known mathematicians as Hadamard, Cartan, Esclangon and the initiator Van der Pol conference. From the USSR, Academicians L. I. Mandelstam, N. M. Kryloff and N. D. Papaleksi, author of this article, are invited as well as young students of the Mandelstam's Academy: Andronov and Vitt.''

\end{quote}

Those who emerge as the world's greatest specialists in the field of nonlinear oscillations were invited at the initiative of Van der Pol, whose unifying role will be highlighted. Unfortunately, due to the flu epidemic raging at that time in Europe, much of those invited did not come. This conference was finally held in the presence of\\

\begin{itemize}
\item Balthazar Van der Pol (Pays-Bas),
\item Alfred Li\'{e}nard (1869-1958), \'{E}lie Cartan (1869-1951) and Henri Cartan (1904-2008), Ernest Esclangon (1876-1954), Henri Abraham (1868-1943), L\'{e}on Brillouin (1889-1969), Philippe Le Corbeiller (1891-1980), Yves Rocard (1903-1992) and Camille Gutton (1872-1963) (France),
\item Charles Manneback (Belgique) and
\item Nikola\"{\i} Papaleksi (URSS).\\
\end{itemize}

According to Papaleksi \cite[p. 210]{Pa1}, Van der Pol would have chosen himself the participants:

\begin{quote}

``The aim of this international conference on nonlinear oscillations (which is the subject of this article and which was convened at the initiative of one of the pioneers of the domain, prof. Van der Pol) was to bring together people from different countries working in this area and give them the opportunity to discuss and exchange views on problems of nonlinear oscillations, to establish a common terminology and at least partially define the direction of future research.''

\end{quote}

Even reduced, this list comprises a major part of the French scientific community strongly involved in the field of nonlinear oscillations. The Belgian Manneback Charles (1894-1975) was in fact a specialist of the  electromagnetism and of the electrodynamics. In a biographical note on Charles Manneback, Mark de Hemptine and Maurice A. Biot mention a stay at the I.H.P.

\begin{quote}

``His teaching qualities highly appreciated abroad earned him many invitations in both Europe and America, and include, among others in 1932-1933 an invitation for a series of lectures at the Institut Poincar\'{e}.\footnote{See
\textit{Annuaire de l'Acad\'{e}mie Royale de Belgique} (1978), p. 14.}''

\end{quote}

In the first part of his article Papaleksi \cite[p. 209-210]{Pa1} pointed out, the importance of process of nonlinear oscillations in theoretical and applied scientific fields such as physics, mechanics, acoustics, biology and radio engineering. Then, he recalled that a first conference of this type was held in Moscow in November 1931 and dispelled thus the confusion between the First All-Union Conference on Oscillations and the International Conference on Nonlinear Oscillations which took place in Paris in January 1933 (See Tab. 1). Nevertheless, it is important to notice that neither the Moscow conference nor the Paris conference should be considered as ``International'' since only members of the Soviet scientific community attended to the former while the latter, due to a flu epidemic, could count among his guests only a majority of French scientists: one Belgian, one Dutch and one Russian.

\section{The Conference Proceedings}

The article of Papaleksi \cite{Pa1} provides the report of discussions and debates that took place during this conference. As the organizer, Van der Pol gave an opening address in which he regretted that many members were absent due to a flu epidemic then raging in Europe. The first speaker was Philippe Le Corbeiller.

According to Papaleksi \cite[p. 210]{Pa1}, Le Corbeiller has presented different types of self-oscillating systems (self-sustained systems) such as the ``metaphorical'' example of ``vase of Tantalus'' that he had introduced during his lectures at the \textit{Conservatoire National des Arts et M\'{e}tiers} (C.N.A.M.) and such as the neon lamp \cite[p. 6]{LeCorb2} he had used to illustrate the relaxation phenomenon. He added that Le Corbeiller has also included in this list of relaxation oscillators the fluctuations of animal species in the struggle for existence of Volterra \cite{Vol1,Vol2,Vol3}. This is somewhat surprising because the predator-prey model of Volterra does not exhibit any relaxation oscillations in its original conception. Thus, it seems unlikely that Le Corbeiller could have mentioned the work of Volterra. However, Papaleksi makes explicitly reference to the book entitled ``Le\c{c}ons sur la theorie
math\'{e}matique de la lutte pour l'existence'' of Volterra \cite{Vol3} who was supposed to be present and thus takes up the example proposed by Andronov \cite[p. 560]{Andro2} in his note at the C.R.A.S. He then stated that Le Corbeiller has reviewed the various problems encountered in the theoretical study of self-sustained oscillations as the condition of existence of a stationary amplitude regime and has also recalled the period value and the form of relaxation oscillations. Papaleksi \cite[p. 210]{Pa1} said that Le Corbeiller has also made a brief historical presentation (probably the same he had already done during his presentation at C.N.A.M.):

\begin{quote}

``After recalling that the linear methods can no longer be used to analyze these problems, he made a small summary of the historical evolution of nonlinear mathematical theory. He then briefly describes the major work of Van der Pol concerning the oscillating systems of Thomson type, then he focused on the work of Cartan and Li\'{e}nard and referred to the link between the graphical analysis of Van der Pol, the geometric construction of M. Li\'{e}nard and the theory of limit cycles of Poincar\'{e} whose importance to the problems of oscillations has been noticed by our young scientists: A. Andronov and A. Witt. Turning to the relaxation oscillations, the speaker mentioned the role played by the research of Van der Pol in this area and indicated that the deepening of the theory of these oscillations has been done by scientists from the USSR.''

\end{quote}

From January 1933, the most representative members of the French scientific community working or having worked in the field of nonlinear oscillations could no longer ignore the ``correspondence'' established by Andronov \cite{Andro1,Andro2} between the periodic solution of an oscillator of Van der Pol's type and the theory of limit cycles introduced by Poincar\'{e} in his memoirs \cite{P2}. Among Cartan, Li\'{e}nard, Abraham and Gutton, none has made use of this result. For most of them this type of research was not at the center of their concerns. As for \'{E}lie and Henri Cartan for example, it is their one and only one article on this subject. Moreover, excepted Henri Cartan, Brillouin, Le Corbeiller and Rocard all other French scientists present were nearly retired (Li\'{e}nard (64 years), \'{E}lie Cartan (64), Esclangon (57), Abraham (65) , Gutton (61 years)). Papaleksi \cite[p. 210]{Pa1} then explained:

\begin{quote}

``In the remainder of his presentation Le Corbeiller focused on the field of quasi-periodic oscillations. He gave a summary of the work of Van der Pol on addressing the problem of forcing periodic self-oscillating systems and stopped on the phenomena of forced synchronization (or ``mitnehmen'') and stressed the importance of these phenomena from a theoretical and experimental point view. Pointing out the importance of the rigorous analysis of synchronization phenomena that was made by A. Andronov and A. Witt, he finished his speech by recalling that new research done in the field of periodic forcing of nonlinear systems and phenomena of fractional resonances were to be presented in the talk of the representative of the USSR.''

\end{quote}

Van der Pol then presented a historical summary of his own work without detailing them but by simply indicating that the references were in the special volume of the \textit{Zeitschrift f\"{u}r Technische Physik} of the USSR, Vol. 4, n$^o$ 1. After his presentation, the subject of discussions was the phenomenon of forced synchronization. According to Papaleksi \cite[p. 211]{Pa1} the next talk was that of Li\'{e}nard in which he recalled the main results of his study on maintained oscillations \cite{Lien1}.

\begin{quote}

``Starting from its graphical method for constructing integral curves of differential equations, he deduced the conditions that must satisfy the nonlinear characteristic of the system in order to have periodic oscillations, that is to say for that the integral curve to be a closed curve, i.e. a limit cycle.''

\end{quote}

This statement on Li\'{e}nard must be considered with great caution. Indeed, one must keep in mind that the narrator has an excellent understanding of the work of Andronov \cite{Andro1,Andro2} and that his report is also intended for members of the Academy of the USSR to which he must justify his presence in France at this conference and show the important diffusion of the Soviet work in Europe. The case of Li\'{e}nard is very special. Indeed, after demonstrating in 1928 in a paper entitled: ``\'{E}tude des oscillations entretenues\footnote{``Study of sustained oscillations''}'' the existence and uniqueness of the periodic solution of the generalized Van der Pol's equation he participated to the \textit{Third International Congress for Applied Mechanics} at Stockholm with Le Corbeiller and Van der Pol. Li\'{e}nard \cite{Lien2} submitted there an article entitled: ``Oscillations auto-entretenues\footnote{``Self-sustained oscillations''}'' in which he generalized a result of Andronov on the stability of periodic solution and in which he made explicitly reference to the note of Andronov \cite{Andro2} of 1929 but still did not speak of Poincar\'{e}'s limit cycle although he used to state his proof the Poincar\'{e}'s method of calculus of variations. Moreover, in the pagination of the Proceedings, the article of Li\'{e}nard \cite[p. 173-177]{Lien2} exactly precedes that of Van der Pol \cite[p. 178-180]{VdP5} in which he also recalled the ``correspondence'' established by Andronov \cite{Andro1,Andro2}.\\

On Sunday, January 29, 1933, astronomer Ernest Esclangon presents his work on the quasi-periodic functions and highlights their importance in particular for the treatment of statistical data. Papaleksi \cite[p. 211]{Pa1} then explains:

\begin{quote}

``During the discussions that took place after the speech, researchers have considered the possibility of observing states of quasi-periodic oscillations in conservative and non conservative nonlinear oscillating systems.''

\end{quote}

The Monday 30 January is dedicated to visit the laboratory of Abraham at the \'{E}cole Normale Sup\'{e}rieure. The various experiments whose Abraham has spoken during his talk are presented as well as the work done in collaboration with Eug\`{e}ne Bloch (may be the ``multivibrator''), including the calibration of the frequency of an oscillator stabilized by a quartz lamp. After this visit, Van der Pol delivers a closing speech, recalling that despite the absence of many eminent scientists in this field, the conference achieved its main purpose to bring together physicists and mathematicians from different countries working in the field of nonlinear oscillations. Then, according to Papaleksi \cite[p. 212]{Pa1}, he added:

\begin{quote}

``Since this first experiment was successful it is hoped that such meetings will be repeated regularly. The main presentations of the Conference should be published\footnote{This publication has obviously not occurred.} in the collection ``Actualit\'{e}s Scientifiques et Industrielles'' (\'{e}dition Hermann, Paris).''

\end{quote}

\section{Discussion}

Because of the absence of many personalities, this conference has not had the international resonance expected by its organizers. It did not lead to the creation of a nonlinear community desired by Van der Pol and it did not lead either to the emergence in France of a \textit{Theory of Oscillations} expected by Le Corbeiller. Nevertheless, although the French mathematicians, physicists and engineers such as Abraham \& Bloch, Blondel, Cartan and Li\'{e}nard have provided major and fundamental contributions for the development of such a Theory they didn't pursued their research in this domain. Among those who continued their work in this area, as Le Corbeiller and Rocard, no form of collaboration with members of the Soviet scientific community (or with Van der Pol also) seemed to have occurred. Their work in this field do not seem to have been influenced significantly by the various presentations including that of Papaleksi or discussions that followed. Thus, although the absence of mathematicians such as Hadamard and Volterra, the most likely to appreciate the work of Poincar\'{e}, was certainly very damaging as well as that of Mandel'shtam and his students Andronov and Witt and that of Kryloff and Bogoliouboff, this first \textit{International Conference on Nonlinear Oscillations} doesn't seem to have produced any impact on the French scientific community. The next conference on \textit{Nonlinear Vibrations} will be held in France between 18 and 21 September 1951 on the island of Porquerolles (Var) in the presence of Van der Pol.

\begin{table}[htbp]
\begin{center}
\tbl{International Conference on Nonlinear Oscillations (I.C.N.O.)}
{\begin{tabular}{|c|c|c|}
\hline
\multicolumn{3}{|c|}{\par International Conference on Nonlinear Oscillations (I.C.N.O.) \par }  \\
\hline
\hspace{1.45cm} Edition \hspace{1.45cm} & \hspace{1.45cm} Year \hspace{1.45cm} & \hspace{1.45cm} Location \hspace{1.45cm} \\
\hline
I$^{st}$ & 1961 & Kiew \\
\hline
II$^{nd}$ & 1962 & Warsow \\
\hline
III$^{rd}$ & 1964 & Berlin \\
\hline
IV$^{th}$ & 1967 & Prague \\
\hline
V$^{th}$ & 1969 & Kiev \\
\hline
VI$^{th}$ & 1972 & Poznan \\
\hline
VII$^{th}$ & 1975 & Berlin \\
\hline
VIII$^{th}$ & 1978 & Prague \\
\hline
IX$^{th}$ & 1981 & Kiew \\
\hline
X$^{th}$ & 1984 & Varna \\
\hline
\end{tabular}}
\end{center}
\end{table}

\nonumsection{Acknowledgments} \noindent

Author would like to thank Pr. Giuseppe Rega, from University of Roma ``La Sapienza'', for his helpful advices.

\end{document}